\documentclass[amsmath,showpacs,twocolumn,aps,prb]{revtex4}
\usepackage{bm}
\usepackage{graphics}
\usepackage{graphicx}

\newcommand{\bq}{\begin{equation}}
\newcommand{\ee}{\end{equation}}

\usepackage[dvips]{color}

\begin{document}

\title{One-dimensional plasmons confined in bilayer graphene $p$-$n$  junctions}
\author{N.~M.~Hassan}
\author{V.~V.~Mkhitaryan}
\author{E.~G.~Mishchenko}
\affiliation{Department of Physics and Astronomy, University of
Utah, Salt Lake City, Utah 84112, USA}


\begin{abstract}
Gapless spectrum of graphene allows easy spatial separation of
electrons and holes with an external in-plane electric field. Guided
collective plasmon modes can propagate along the separation line,
with the amplitude decaying with the distance to it. Their spectrum
and direction of propagation can be controlled with the strength and
direction of in-plane electric field. In the case of a bilayer graphene
additional control is possible by the perpendicular electric field
that opens a gap in the band spectrum of electrons. We investigate
guided plasmon spectra in bilayer {\it p}-{\it n} junctions using
hydrodynamics of charged electron liquid.

\end{abstract}

\maketitle

\section{Introduction}

Atomically thick graphene\cite{Wil} allows control of its electrical
properties in a variety of ways and thus is of great potential for
nanoelectronics\cite{GN} and optoelectronics.\cite{HAA,SAM} One of
the major advantages lies with the gapless nature of its electron
band spectrum, which means that relatively moderate fields are
required to induce desirable changes in the density of electrons and
holes. This is in a sharp contrast to conventional nanoplasmonics of metal particles whose properties are fixed for once at synthesis.
Plasmons in graphene, due to much higher tunability of the latter,
acquire new properties. While the plasmon spectrum in a homogeneous
graphene sheet subject to finite temperature \cite{T_plasmon} or
doping \cite{E_plasmon} follows the conventional two-dimensional (2D)
form,\cite{Stern} $\omega^2(q)\propto q$ , spatial separation of
electrons and holes leads to novel excitations. In particular,
one-dimensional plasmons can propagate along a graphene {\it p-n}
junction, \cite{MSS} with the spectrum $\omega^2(q)\propto
(E_0q)^{1/2}$, which depends on the strength of electric field $E_0$ that
separates electrons and holes. Thus both the direction of
propagation and velocity (for a given wavelength) of these
{\it guided} plasmons can potentially be controlled by changing the
orientation and magnitude of the electric field that creates the
{\it p-n} junction.

Bilayer graphene\cite{cn} is potentially of even  greater
significance for applications because of the possibility of opening
a bandgap via external gating,\cite{MCF} and thus transitioning from
the metallic to the insulating state.  Correspondingly, plasmon
excitations in bilayer graphene are a subject of active
research. \cite{WC,SHD,JBS,G} However, to date, only 2D plasmons in
homogeneous bilayers have been studied. In the present paper we
consider guided one-dimensional plasmons propagating along {\it p-n}
junctions in bilayer graphene, in an extension of the work in
Ref.~\onlinecite{MSS}. Bilayer {\it p-n} junctions have recently
been reported experimentally \cite{JJ} and their transport
properties are attracting increased theoretical interest.
\cite{P,NL,PS} Here we address their collective modes.

Our objective is to study the behavior of
one-dimensional plasmons propagating in a bilayer graphene and the
dependence of the plasmon spectrum on the electric field $E_0$. In
this work we consider two cases; the first case is plasmon
propagation in a bilayer graphene flake under the effect of only the
in-plane electric field $E_0$ (gapless case), discussed in Secs.~II and III. In the second case, discussed in Sec.~IV, there is an
additional electric field perpendicular to the plane of graphene
that opens a gap in the band electron spectrum. In both cases the
system is a flake of width $2 d$ (in the $x$ direction) and infinite
length along the $y$ axis, as shown in Fig.~1(a). The in-plane electric
field ${\bf E_0}$ is applied along the x axis. Due to the gapless
nature of the spectrum of bilayer graphene, electric field induces
charge separation and produces charge density $\rho_0(x)$ across the
width of the flake such that one half of the flake is positively
charged and the other half is negatively charged. Fluctuations of
charge density $\delta\rho$ propagate on top of the equilibrium
density in the form of plasmon waves. Because of the nonuniform
profile of $\rho_0(x)$  plasmons are localized near the neutrality
line $x=0$ and behave as quasi-one-dimensional excitations. We
utilize the Thomas-Fermi equation to find the equilibrium charge
density $\rho_0$ and then use hydrodynamic equations to describe the
dynamics of charge oscillations $\delta\rho({\bf r},t)$. It reduces
to solving an integrodifferential eigenvalue problem for the
plasmon frequencies. For the ungated case in the limit of short
wavelengths $\ll d$ even eigenfrequencies have higher values than
the odd ones. Interestingly, at large wavelengths $\gg d$ this order
is changed and the lowest eigenmode is an even one. In all cases the
plasmon frequencies are proportional to $\sqrt{E_0}$. In the gated
case, the perpendicular electric field and the ensuing energy gap
leads to the appearance of a  neutral strip at the center of the
flake. As a result the spectrum becomes linear in $E_0$ signifying
the increased sensitivity to the applied field in a gated bilayer.

\section{Hydrodynamics of a bilayer}

In this section we describe the formalism and
derive the plasmon spectrum in a gapless bilayer. In part A the
equilibrium charge density $\rho_0(x)$ is derived within the
Thomas-Fermi approximation. In part B we obtain an
integrodifferential equation (\ref{eq30}) for  $\delta\rho({\bf
r},t)$ and solve it in the short-wavelength limit. The solutions
have a pseudocontinuous spectrum Eqs.~(\ref{evenspec}) and (\ref{oddspec}),
which becomes discrete, see Eq.~(\ref{dispersion}) and Eq.~(\ref{ns}), after
logarithmic singularities are regularized. The limit of long
wavelengths is discussed in part C and a surprising minimum in the
frequency of odd modes is noticed at intermediate wavelengths $\sim
d$. This behavior is further discussed in part D.

We begin with a description of the mean
induced charge profile $\rho_0(x)$ and fluctuation $\delta\rho({\bf
r},t)$. The applied electric field induces charge density of
electrons (or holes) $\rho_0(x)$, which has to be found by taking the
electric field of the induced charges into account
self-consistently. The induced carrier density is easily estimated
by the order of magnitude. Consider that the typical distance
between the charges is denoted by $l_E$. The equilibrium is reached
when the amount of induced charge becomes sufficient to balance the
applied external field, $e/l_E^2 \approx E_0$, which implies for the
average density, $\overline{|\rho_0|} \sim e/l_E^2 \sim E_0$. In the
case when $l_E\ll d$ there are many induced charges across the width
of the flake and a continuous description can be used. In order to
apply the semiclassical approximation one more condition has to be
satisfied. In particular, the plasmon wavelength $\lambda$ has to
be much greater than the Fermi wavelength of the induced carriers.
The latter can be estimated from $k_F^2 \sim \overline{|\rho_0|}/e
\sim 1/l_E^2$. We conclude that it is necessary to ensure that
\begin{equation}
\label{lE} l_E\equiv \sqrt{\frac{e}{E_0}} \ll d,~\lambda.
\end{equation}
\begin{figure}
\label{fig1}
\resizebox{.38\textwidth}{!}{\includegraphics{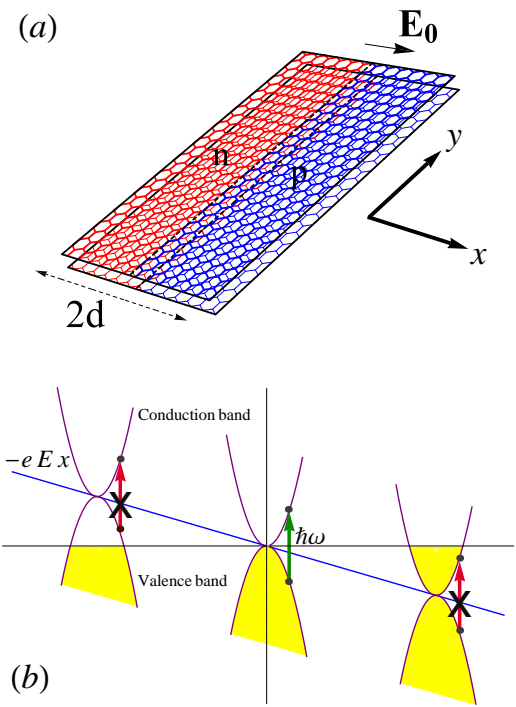}}
\caption{(Color online) (a) Bilayer graphene flake of width $2d$ is placed in
electric field that separates {\it p} and {\it n} regions. Guided
plasmons propagate along the $y$ axis and decay in the
$x$ direction. (b) Schematic picture of the electron band structure:
in equilibrium the sum of the electrostatic potential and kinetic
energy of electrons (at the Fermi level) is constant. Vertical
arrows indicate the possibility of optical absorption with a given
frequency: while possible near the center of the flake (green non-crossed
arrow), such transitions are forbidden (red crossed arrows) when both
initial and final states are empty or occupied.}
\end{figure}

Under these conditions macroscopic hydrodynamic equations of the
charged liquid can be applied. For the total electric field we have
\begin{equation}
\label{eq19} {\bf E}({\bf r},t) ={\bf E}_0-\nabla \int d^2r' \frac{\rho ({\bf r'},t)}{|{\bf r}-{\bf r'}|}.
\end{equation}
The fluctuating density $\rho ({\bf r},t)$ obeys the  charge
conservation law
\begin{equation}
\label{eq18} \dot \rho ({\bf r},t) + \nabla \cdot {\bf
J}({\bf r},t)=0.
\end{equation}
While the two relations (\ref{eq19}) and (\ref{eq18}) are quite generic,
the remaining equation for the electric current is system specific.
For single-layer graphene it has been discussed
in Refs.~\onlinecite{sachdev} and \onlinecite{MSS}. The band structure of bilayer graphene consists of
four bands that originate from the coupling of two Dirac cones of the
two layers.\cite{cn} The two outside bands are separated by a large
band gap $\sim0.6$ eV and will therefore be ignored here. The
remaining two bands touch each other (in the absence of interlayer
bias) and are parabolic, $\pm p^2/2m$, see Fig.~1, with the
effective mass $m=0.05m_0$, where $m_0$ is the vacuum electron mass.
For parabolic bands we find (see Appendix A for derivation)
\begin{equation}
\label{eq17} {\bf \dot J}=\frac{|\rho|}{m}\left(e{\bf
E}-\nabla \mu  \right);
\end{equation}
here the chemical potential is related to the induced charge density
via $\mu =-\pi\rho/2me$, where the fourfold (spin and valley)
degeneracy is taken into account. The form of Eq.~(\ref{eq17}) is
readily recognized from the usual ac Drude conductivity $\sigma={i e
|\rho|}/{m\omega}$ if one notes that the expression in the brackets
is simply the gradient of the electrochemical potential. The main
difference comes from the fact that the effective ``Drude
conductivity'' can depend on the coordinates (and time) via
$|\rho|$.

\subsection{Mean charge distribution}

In equilibrium electric current is absent, ${\bf J}=0$, and the mean
density of charges $\rho_0(x)$ obeys the equation
\begin{equation}
\label{TFe} {E_0x+2 \int_0^d dx' \rho_0(x')
\ln{\frac{x+x'}{|x-x'|}}+\frac{\pi a_B}{2} \rho_0(x) =0}.
\end{equation}
This is the Thomas-Fermi equation where the first two terms
describe electrostatics of an ideal metal strip. The last term
takes into account that the screening radius is finite, the latter
being  of the same order as the Bohr radius $a_B={\hbar^2}/{m
e^2}\approx 10.6 ~\AA$. When the width of the flake is much
greater than the Bohr radius, $d\gg a_B$, which is the case for any practical
situation, the last term in Eq.~(\ref{TFe}) is negligible and the
solution to the remaining integral equation is simply
\begin{eqnarray}
\label{eq30-1} \rho_0(x)=\frac{E_0x}{\sqrt{d^2-x^2}}.
\end{eqnarray}
This solution is known \cite{SMR} from the method of conformal
mapping for the two-dimensional Laplace equation, but it is most
easily verified by a direct substitution into Eq.~(\ref{TFe}). Expression (\ref{eq30-1}) is applicable everywhere except very
close to an edge,  where in principle the last term in
Eq.~(\ref{TFe}) would provide regularization of the square-root
singularity in the mean density. Such a procedure, however,
strictly speaking, would exceed the accuracy of our semiclassical
treatment. Indeed, as follows from Eq.~(\ref{eq30-1}), the last
term in Eq.~(\ref{TFe}) becomes comparable with the first two at
$d-x \sim a_B^2/d$, which is a length of the order of lattice
spacing (or even smaller), where fully quantum-mechanical
treatment is warranted. Fortunately for our purposes, the
singularity in Eq.~(\ref{eq30-1}) is integrable and its presence
will not affect the subsequent calculations.

\subsection{Plasma oscillations}

Plasmons are charge density oscillations propagating on top of the
equilibrium profile , Eq. (\ref{eq30-1}),
\begin{equation}
\rho=\rho_0(x)+\delta \rho({\bf r},t).
\end{equation}
Equations (\ref{eq19})-(\ref{eq17}) can be linearized with respect
to the density  variation $\delta \rho({\bf r},t)$. The latter will
be taken in the form of a plane wave propagating along the junction,
$\delta \rho({\bf r},t)=\chi(x) \exp(i\omega t - iqy)$. By substituting for
electric field ${\bf E}$ and current ${\bf J}$ in Eq.~(\ref{eq18}) we obtain the
following integrodifferential equation for the oscillating density
profile,
\begin{eqnarray}
\label{eq30} \omega^2 \chi(x)+\frac{2 e}{m}\left[ \frac
{d}{d x}|\rho_0(x)|\frac {d}{d x} -q^2|\rho_0(x)|\right] \nonumber\\
\times \int_{-d}^d dx'\chi(x') K_0(q|x-x'|)=0,
\end{eqnarray}
where $K_0$ is the modified Bessel function of the second kind.
Note that in deriving Eq.~(\ref{eq30}) we neglect the $\nabla \mu$
term in Eq.~(\ref{eq17}). As  can be readily seen, its contribution
is small by the parameter $a_B/\lambda \ll 1$.

Equation (\ref{eq30}) is reminiscent of the equation for
single-layer graphene,\cite{MSS} where $\sqrt{|\rho_0(x)|}$ takes
the place of $|\rho_0(x)|$. This difference makes the solution of the
bilayer problem both simpler and trickier. In the most interesting
case of a wide strip, $\lambda \ll d$, the confinement of plasmons
originates from the {\it gradient} of the equilibrium charge
density. Since, as we find below, plasmon oscillations extend over
distances of the order of their wavelength in the transverse
$x$ direction, the limits of integration in Eq.~(\ref{eq30}) can be
extended to infinity while the mean density approximated with
$\rho_0(x)=E_0x/d$. The plasmon momentum can then be conveniently
scaled away with the substitution $qx=\xi$. Furthermore, using the
equation for the modified Bessel function, $\tau K_0''(|\tau|)
+K_0'(|\tau|)=\tau K_0(|\tau|)$, it is convenient to rewrite the
differential operation in Eq.~(\ref{eq30}) as follows:
$(\frac{d}{d\xi}|\xi|\frac{d}{d\xi}-|\xi|)K_0(|\xi-\xi'|)=\xi'\text{sgn}~\xi(\frac{d^2}{d\xi^2}-1)K_0(|\xi-\xi'|)$. Equation (\ref{eq30})
then becomes ($\xi=qx$):
\begin{eqnarray}
\label{integral_eq1} \omega^2 \chi(\xi)+\frac{2 eE}{md}\text{sgn}~\xi
 \biggl(\frac{d^2}{d\xi^2}-1\biggr) \nonumber\\ \times \int_{-\infty}^\infty d\xi'
\xi'\chi(\xi') K_0(|\xi-\xi'|)=0,
\end{eqnarray}
that in the Fourier representation acquires the form
\begin{eqnarray}
\label{eq53} \omega^2 \chi(k)=-\frac{4\pi e E_0}{md} ~{P}
\int_{-\infty}^\infty \frac{dk'}{2\pi}\,
\frac{\sqrt{1+k'^2}}{k'-k}\frac{d\chi(k')}{dk'},
\end{eqnarray}
where $P$ stands for the principal value of integral. The integral
equation (\ref{eq53}) should determine the discrete spectrum of
plasmon eigenvalues $\omega_n$. As expected, from the symmetry of the
system, the solutions possess definite parity. It can be seen that
Eq.~(\ref{eq53}) has even solutions
\begin{eqnarray}
\label{evensol}
\chi^{(+)}_\alpha(k)=\cosh{\left(\frac{\pi\alpha}{2}\right)}
\cos\Bigl(\alpha~\text{arcsinh}~{k}\Bigr),
\end{eqnarray}
and odd solutions
\begin{eqnarray}
\label{oddsol}
\chi^{(-)}_\alpha(k)=i\sinh{\left(\frac{\pi\alpha}{2}\right)}
\sin\Bigl(\alpha~\text{arcsinh}~{k}\Bigr),
\end{eqnarray}
where $\alpha$ is an {\it arbitrary} positive number. This follows
from the following relations
\begin{eqnarray}
\label{intrel1} && {P} \int\limits_{-\infty}^\infty dt
\frac{\cosh t\sin(\alpha t)}{\sinh t-\sinh \tau}=\frac{\pi
\cos(\alpha\tau)}{\tanh(\pi\alpha/2)}
,\\
&&{P}\int\limits_{-\infty}^\infty dt \frac{\cosh t\cos(\alpha
t)}{\sinh t-\sinh \tau}=-\pi\tanh(\pi\alpha/2)
\sin(\alpha\tau),\qquad \label{intrel2}
\end{eqnarray}
that can be proven by calculating the corresponding integrals with
the help of adding an infinite semicircle in the complex plane
and summing over the residues; see Appendix B. The corresponding
spectrum of eigenvalues is gapped for the even modes,
\begin{eqnarray}
\label{evenspec} \omega^2_{+}(\alpha)= \frac{2\pi e
E_0}{md}\frac{\alpha} {\tanh(\pi\alpha/2)},
\end{eqnarray}
and gapless for the odd modes,
\begin{eqnarray}
\label{oddspec} \omega^2_{-}(\alpha)= \frac{2\pi e E_0}{md}\alpha
\tanh(\pi\alpha/2).
\end{eqnarray}
Finally, we note that the even and odd solutions obey a very simple relation in the real space,
\begin{eqnarray}
\label{evenoddrel} \chi^{(-)}_\alpha (\xi)=\text{sgn}(\xi) \chi^{(+)}_\alpha (\xi).
\end{eqnarray}
This can be obtained by noticing that $\chi^{(+)}(k)$ and
$\chi^{(-)}(k)$ are related by the Hilbert transform: the direct
substitution of the solutions (\ref{evensol}) and (\ref{oddsol})
into Eq.~(\ref{eq53}) yields
\begin{equation}
\chi^{(+)}_\alpha (k)=-i {P}
\int_{-\infty}^\infty \frac{dk'}{\pi}\frac{\chi^{(-)}_\alpha (k')}{k'-k},
\end{equation}
which is the Fourier transform of Eq.~(\ref{evenoddrel}).

The most surprising feature of the obtained solutions is the
continuity of the spectrum of plasmon frequencies, which is in a
seeming contradiction to the fact that plasmon modes are localized
in the transverse direction. To elucidate the physical origin of
this finding let us find the explicit form of our solutions in real
space. Due to Eq. (\ref{evenoddrel}) it is sufficient to consider
$\chi_\alpha^{(+)}(\xi)$ for positive arguments, $\xi>0$. As shown
in Appendix B the solution is given by the modified Bessel
function of the imaginary order,
\begin{eqnarray}
\label{Fourier} \chi^{(+)}_\alpha(\xi)=\int\limits_{-\infty}^\infty
\frac{dk}{2\pi} e^{ik\xi} \chi^{(+)}_\alpha
(k)=\frac{\alpha\sinh(\pi\alpha)} {2\pi \xi}\nonumber\\\times
\int\limits_{0}^\infty dt e^{-\xi \cosh t} \cos(\alpha t)
=\frac{\alpha\sinh(\pi\alpha)} {2\pi \xi}K_{i\alpha}(\xi).
\end{eqnarray}
For large distances, $\xi \gg 1$, the asymptotic behavior is exponential,
\begin{equation}
\chi^{(+)}_\alpha(\xi)\propto \frac{e^{-\xi}}{\xi^{3/2}}.
\end{equation}
We indeed obtain that guided plasmons are localized on the
distances of order of their wavelengths $\sim \lambda$.

The behavior at small distances, $\xi \ll 1$, is more peculiar:
\begin{equation}
\label{smallx} \chi^{(+)}_\alpha(\xi)\approx
-\frac{\alpha}{2\xi}\, \text{Im}~
\frac{(\xi/2)^{i\alpha}}{\Gamma(1+i\alpha)}.
\end{equation}
We observe that at small distances the solutions Eq. (\ref{Fourier})
have infinitely many nodes, which is formally responsible for the
continuity of their spectrum. Noteworthy is the similarity between
our charge density $\chi^{(+)}_\alpha$ and the wave function of a
quantum-mechanical particle moving in the attractive potential
$V(\xi)=-{\alpha^2}/{\xi^2}$ in two dimensions, the situation that
leads to a particle {\it falling} on the attraction
center.\cite{LLIII} It is therefore obvious that oscillatory
behavior at $\xi\to 0$ is an artifact of the semiclassical
approximation. The latter fails at small distances comparable with
the Fermi wavelength (which gets large closer to the {\it p-n}
junction line). Thus such fast oscillations are unphysical and would
not have appeared in the fully quantum-mechanical treatment of
electrons in electric field. Fortunately, the oscillatory behavior
is only logarithmic and can be easily regularized. This can be
performed by noting that  the solutions $\chi_\alpha ^+(\xi)$ {\it
smoothed} over these fast unphysical oscillations vanish at $\xi \to
0$. The vanishing is most easily seen from the density accumulated
at small distances: using Eq.~(\ref{smallx}) we find
\begin{equation}
\label{net_charge} \int\limits_0^\xi d\xi' \chi^{(+)}_\alpha(\xi')
=A\cos{[\alpha \ln{(\xi/2)}-\beta]},
\end{equation}
where $\beta$ is the phase of the complex quantity
$\Gamma(1+i\alpha)$. The  accumulated charge therefore oscillates
with constant amplitude $A$ around zero value. We therefore impose
the regularization requirement that the smoothed solution should
vanish at distances smaller than some  $a$  (see below) where the
semiclassical approach ceases to be valid. From Eq.~(\ref{Fourier})
we obtain the equation
\begin{equation}
\label{dispersion}
K_{i\alpha}(qa)=0,
\end{equation}
which determines the set of discreet values of $\alpha$.
Note that the choice of Eq.~(\ref{dispersion}) is
natural for both the even and odd modes as the replacement of the
oscillatory function $K_{i\alpha}(\xi)$ with any smoothed nonzero
value would have led to the nonintegrable singularity in Eq.
(\ref{Fourier}) at $\xi=0$.

\begin{figure}[t]\vspace{-0.4cm}
\resizebox{0.5\textwidth}{!}{\includegraphics{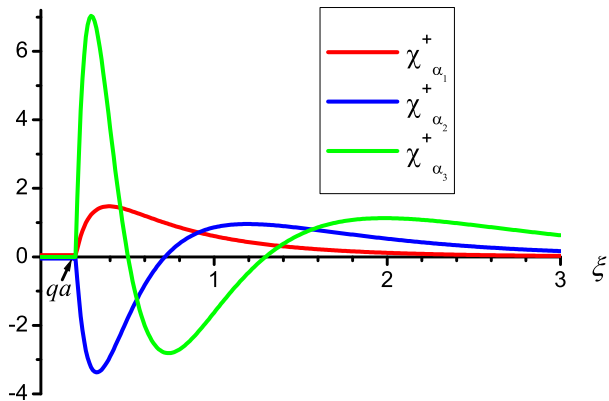}}
\caption{\label{profiles} (Color online) The eigenfunctions
$\chi^{(+)}_{\alpha_n}(\xi)$, Eq. (\ref{Fourier}), for three lowest
values $n=1,\,2$, and $3$ and $qa=0.2$. The first mode (red) is node-less while the second one (blue) has just one node and the third (green) has two nodes and the sharpest peak. For the sake of convenience
the functions $2 \chi^{(+)}_\alpha(\xi)/\cosh{(\pi\alpha/2)}$ are
plotted.  The corresponding eigenvalues for $\alpha_1=1.44$,
$\alpha_2=2.49$, $\alpha_3=3.42$ are found from
Eq.~(\ref{evenspec}).}
\end{figure}

Equation (\ref{dispersion}) constitutes the quantization
condition that together with Eqs.~(\ref{evensol}) and
(\ref{oddsol}) determines the spectrum of guided plasmon modes for
 wavelengths shorter than the width of the flake. For $\alpha
\ll 1$ we obtain in the logarithmic accuracy the following analytic approximation:
\begin{equation}
\label{ns}
\alpha_n=\frac{\pi n}{\ln{\displaystyle \left(\frac{2e^{-\gamma}}{qa}\right)}}, ~~~n=1,2,3,\,.\,.\,.\,,
\end{equation}
where $\gamma=0.58$ is the Euler's constant.  The first three eigenfunctions
$\chi^{(+)}_{\alpha_n}(\xi)$, $n=1,\,2$, and $3$, are plotted in
Fig.~\ref{profiles}.

The cutoff distance $a$ can be estimated as follows. The
semiclassics fail at the distances of the order of the Fermi
wavelength, $a\sim k_F^{-1}$. The latter, however, is a function of
the coordinate, $k_F\sim (E_0 x/d e)^{1/2}$, cf.~Eq.~(\ref{eq30-1}),
and should be taken at the same distance, $x\sim a$. We therefore
obtain that $a\sim l_E (d/l_E)^{1/3}$, i.e., that the cutoff is
mostly given by the electric length defined in Eq.~(\ref{lE}) and
exceeds the latter only by virtue of the factor $(d/l_E)^{1/3}$.

\subsection{Long wavelengths, $q\ll 1/d$}

When the plasmon wavelength becomes comparable with $d$ (or exceeds
it) the oscillating electric field extends beyond the width of the
flake. The integral equation (\ref{eq30}) has to be solved with the
explicit boundary condition (unimportant previously) requiring that
there is no particle flow across the edges of the system,
\begin{equation}
\label{add_boundary} J_x(\pm d)=0.
\end{equation}
In the limit $q\to 0$ the properties of plasmon spectrum can be
elucidated without the exact solution. Eq.~(\ref{eq30}) becomes
\begin{eqnarray}
\label{zeroq} \omega^2(0) \chi(\zeta)=-\frac{2 eE_0}{md} \frac {d}{d
\zeta}\frac{|\zeta|}{\sqrt{1-\zeta^2}} \int\limits_{-1}^1 d\zeta'
\frac{\chi(\zeta')}{\zeta'-\zeta},
\end{eqnarray}
where $\zeta=x/d$. This equation has one zero eigenvalue,
$\omega_1(0)=0$, which corresponds to the function
\begin{equation}
\label{uniform}
 \chi(\zeta)=\frac{1}{\sqrt{1-\zeta^2}}.
\end{equation}
The solution (\ref{uniform}) simply describes charge distribution in
the equipotentially charged metallic strip. For finite but small $q$
this mode develops into the usual one-dimensional plasmon with the
spectrum (cf. Ref. \onlinecite{MSS} for a similar discussion of the
single-layer case)
\begin{equation}
\label{1dplasmon} \omega_{1+}^2(q) = {\cal C}q^2 \ln{(1/qd)}.
\end{equation}
Here we denote by $\omega_{n\pm}$ the eigenfrequency of an even/odd
mode with $n-1$ nodes across the half width of the flake $(0,d)$. In
the short-wavelength limit it simply corresponds to
$\omega_{\pm}(\alpha_n)$; cf.~Eqs.~(\ref{evenspec}) and (\ref{oddspec}).
The constant ${\cal C}$ is proportional to the total number of
induced charges per unit length along the $y$ direction and can be
most simply found by integrating the original Eq.~(\ref{eq30})
across the width of the graphene strip. Using the approximation
$K_0(q|x-x'|)=-\ln{q|x-x'|}$ and noticing that the term containing
the total derivative vanishes by virtue of the boundary condition
(\ref{add_boundary}), we obtain ${\cal C}=4 eE_0d/m$. The logarithm
in Eq.~(\ref{1dplasmon}) originates from the long-range nature of
Coulomb interaction.

Naively, one would expect the lowest-lying odd plasmon solution to
be $(1-)$, i.e., to change its sign once, at the center of the flake,
$\zeta=0$, but to keep the same sign across the {\it half width} of
the sample. Remarkably, Eq.~(\ref{zeroq}) does not admit an odd
solution without at least one zero in the domain $(0,\,d)$, which
also obeys the boundary condition Eq.~(\ref{add_boundary}). To see
this, let us integrate Eq.~(\ref{zeroq}) over $\zeta$ from $0$ to
$1$. The boundary condition (\ref{add_boundary}) implies that, as
$\zeta\rightarrow1$, we get
\begin{equation}\label{fromBC}
J_x(1)\propto \frac{\zeta}{\sqrt{1-\zeta^2}} \int\limits_{-1}^1
d\zeta' \frac{\chi(\zeta')}{\zeta'-\zeta} \Biggl|_{\zeta \to 1}=0.
\end{equation}
Then we arrive at the relation
\begin{equation}\label{intchi}
\frac{\omega^2(0) md}{2 eE_0} \int\limits_{0}^1
\!d\zeta\,\chi(\zeta)=
\lim_{\zeta\rightarrow0}\frac{|\zeta|}{\sqrt{1-\zeta^2}}
\int\limits_{-1}^1 d\zeta' \frac{\chi(\zeta')}{\zeta'-\zeta}.
\end{equation}
In physical terms this means that in order to have a net charge
across the half width, $\int_{0}^1 \!d\zeta\,\chi(\zeta)\neq 0$,
there should be a nonvanishing current across the {\it p-n}
junction, given by the right-hand side in Eq.~(\ref{intchi}).
Nonvanishing of the limit in Eq. (\ref{intchi}) implies in its turn
that at the junction, $\zeta=0$, the induced electric field must be
singularly strong,
\begin{equation}\label{singEx}
E(\zeta\rightarrow0)\propto \int\limits_{-1}^1 d\zeta'
\frac{\chi(\zeta')}{\zeta'-\zeta}\sim\frac 1\zeta.
\end{equation}
To create such a strong field, the plasmon fluctuation should have a
$\delta$-function singularity at $\zeta=0$, $\chi(\zeta)\propto
\delta(\zeta)$. Such a situation is unphysical and in any case in
conflict with the assumption that $\chi(\zeta)$ is an odd function.
Thus we conclude that $\int_0^1d\xi \chi(\xi)=0$, and  the
lowest-lying odd plasmon eigenmode $(2-)$ should have at least three
nodes across the width of the sample rather than a single node as
might be intuitively expected.

The mode (\ref{1dplasmon}) is gapless because its
electric potential is uniform across the flake. All
other modes have nodes and therefore their frequencies do not tend
to zero in the limit $q\to 0$ but instead have energy gaps
$\omega_n(0)$ determined by Eq.~(\ref{zeroq}). Without explicitly
solving it we conclude from scaling that for all plasmons except the
lowest even mode,
\begin{equation}
\label{gaps} \omega^2_n(q\to 0) =\text{const}(n)\times \frac{eE_0}{md}.
\end{equation}
In other words the plasmon spectrum at large wavelengths is
determined by  {\it the same} energy scale $eE_0/md$ as in the case
of short wavelengths (\ref{evenspec}) and (\ref{oddspec}).

\subsection{Mode order reversal at the intermediate wavelengths, $q\sim 1/d$.}

\begin{figure}[t]
\resizebox{.45\textwidth}{!}{\includegraphics{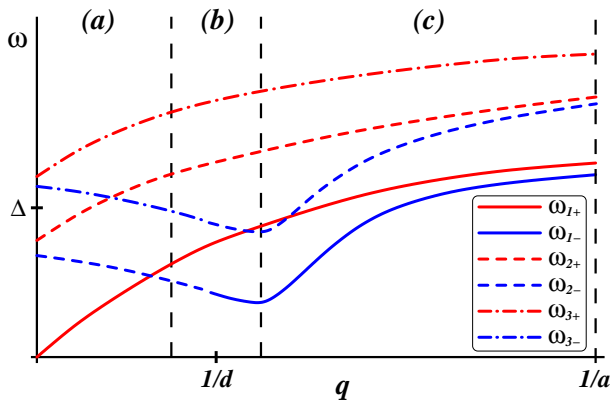}}
\caption{\label{spectrum} (Color online)The sketch of even plasmon (red monotonic curves) and odd
plasmon (blue curves with minima around $1/d$) frequencies for $n=1$ (solid lines) and $n=2$ (dashed
lines). The three regions, $q\ll1/d$, $q\sim1/d$, and $1/d\ll q<1/a$
are denoted by $(a)$, $(b)$, and $(c)$, respectively. With
decreasing the wavelength through $\lambda \sim d$ the $n$th odd
mode at $q\ll 1/d$ develops into the $n-1\,$th odd mode at $q\sim
1/d$, e.g., $(2-)$ becomes $(1-)$.}
\end{figure}

\begin{figure}[h]
\resizebox{.40\textwidth}{!}{\includegraphics{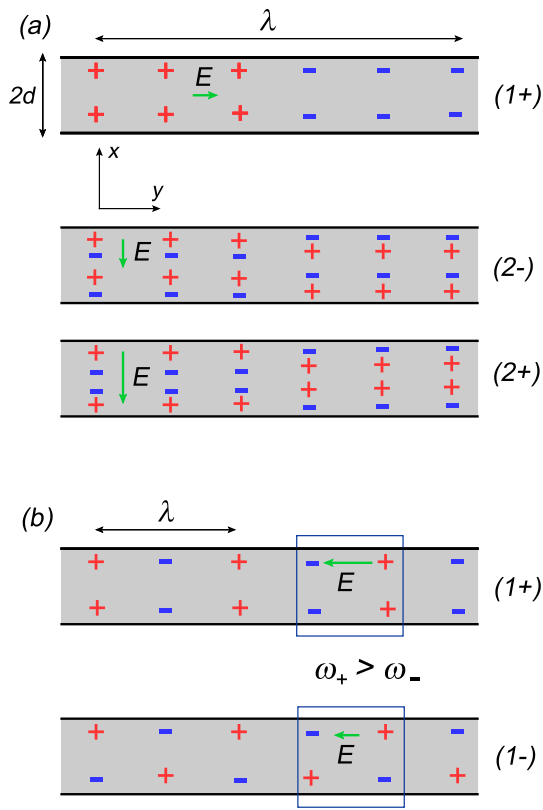}}
\caption{(Color online) Schematic charge distribution for the lowest plasmonic
modes corresponding to different regions in Fig. ~\ref{spectrum}. (a)
For long wavelengths, $\lambda\gg d$, the lowest mode $(1+)$ is the
quasi-one-dimensional plasmon with symmetric nodeless (across the
flake) charge distribution and electric field pointed mostly along
the $y$ direction. The magnitude of electric field vanishes with
increasing the wavelength, in agreement with the gapless spectrum;
Eq.~(\ref{1dplasmon}). The second mode $(2-)$ is odd and gapped and
has three nodes. A mode with a single node $(1-)$ is forbidden by
the condition $\int_{0}^d \!dx\,\chi(x)=0$, valid for any gapped
mode at $q\to 0$. The origin of that condition is the suppression of
charge transport across the line separating $p$ and $n$ regions. The
third mode $(2+)$ is even and has two nodes. The electric field in
the modes $(2-)$ and $(2+)$ is mostly pointed along the
$x$ direction. The  field is stronger for the mode $(2+)$ (this fact
could be easily inferred qualitatively from the picture of charge
distribution), thus leading to $\omega_{2-}<\omega_{2+}$. (b) For
intermediate wavelengths $\lambda\sim d$, the mode $(1-)$ is no
longer forbidden and the $(1-)$ mode at $\lambda \sim d$ evolves
from the $(2-)$ mode at $q=0$ and at $\lambda\sim d$ has the
frequency below that of the $(1+)$ plasmon. This reversal happens
because the checkerboard pattern of the $(1-)$ mode has lower
electric field than the $(1+)$ mode. \label{electrostatics}}
\end{figure}

The results obtained so far have shown surprising reversal in the
order of even and odd solutions. At long wavelengths, $\lambda \gg
d$, shown as domain (a) in Fig.~\ref{spectrum}, the lowest
energy mode is the gapless plasmon $\omega_1(q)$,
Eq.~(\ref{1dplasmon}). However, in the short-wavelength limit,
$\lambda \ll d$, domain (c), the order is reversed as
$\omega_{+}(\alpha_n)>\omega_{-}(\alpha_n)$,
cf.~Eqs.~(\ref{evenspec}) and (\ref{oddspec}). Formally, as the
short-wavelength cutoff is made smaller, $a\to 0$, the values of the
lowest $\alpha_n$ decrease and  the odd frequencies
$\omega_{-}(\alpha_n)$ approach zero; cf.~Eq.~(\ref{oddspec}). Such a
change in the order of even and odd modes  can be explained with
simple physical picture of charge distribution illustrated in
Fig.~\ref{electrostatics} for the first several
modes. The lowest even mode's charge distribution is nodeless across the
width of the flake, and in the case of long wavelength $\lambda\gg
d$, Fig.~\ref{electrostatics}(a), produces a weak restoring
longitudinal field $E \propto 1/\lambda^2$, resulting in the gapless
spectrum (\ref{1dplasmon}). At shorter wavelengths the intermediate domain (b) $\lambda \sim d$, the
distribution of charges for the lowest odd mode resembles a
checkerboard with any plaquette of four charges composing a
quadrupole. In the lowest even mode, on the other hand, the
plaquette consists of two uncompensated dipoles that produce
stronger electric field. We therefore conclude that around $\lambda
\sim d$ the first  odd plasmon becomes the lowest mode of the
system.

For intermediate wavelengths $\lambda\sim d$, the mode $(1-)$ is no
longer forbidden since $\int_{0}^1 \!d\zeta\,\chi(\zeta)\neq 0$, as
the charge transport occurs along the $y$ direction. The $(1-)$ mode
evolves from the $(2-)$ mode at $q=0$ and at $\lambda\sim d$ has the
frequency below that of the $(1+)$ plasmon. This reversal happens
because the ``checkerboard'' pattern of the $(1-)$ mode
significantly reduces the electric field compared with the $(1+)$
arrangement.

It appears that the reason for the change in the
number of nodes in odd solutions is the linear decrease of the
conductivity of a bilayer graphene near the line $x=0$ that prevents
currents from flowing across that line so that the zero-node
profile $(1-)$ can exist only when the wavelength becomes short enough,
$\sim d$, for the longitudinal currents (along the $y$ axis) to be able to
significantly  affect the charge distribution. In
contrast, in a monolayer graphene the conductivity vanishes only as the
square root of $x$ and therefore currents across the junction can potentially remain finite
if the fluctuating field has a $1/\sqrt{|x|}$ singularity, which is the case for odd solutions in a monolayer graphene.\cite{MSS}

Let us emphasize that in plasmonic systems there is no inherent
reason for even/odd modes to always have the same order. In
particular, in 3D metal films the even mode has lower
frequency,\cite{eco} while in a system of two 2D
electron layers separated by a dielectric the situation is
reversed.\cite{dsm} What makes guided plasmons in graphene bilayer
peculiar is the crossover between these two cases for different
wavelengths.

\subsection{Comparison with the case of a
monolayer, Ref.~\onlinecite{MSS}}

It is useful to compare our findings for a bilayer
graphene with the results of Ref.~\onlinecite{MSS} for a monolayer.
In the latter case at short wavelengths $q \gg 1/d$ the solutions were doubly  {\it degenerate}  with a pair of even and
odd modes having the same frequency. The frequencies were
$\propto q^{1/4}$ while in the bilayer case they are [see
Eqs.~(\ref{evenspec}) and (\ref{oddspec})] virtually independent of $q$ [up
to the weak logarithmic dependence in Eq.~(\ref{ns})]. In addition,
the dependence on the strength of the electric field creating the
$p$-$n$ junction was $\propto E_0^{1/4}$ compared with the
$E_0^{1/2}$ dependence for a bilayer.

The above-mentioned monolayer degeneracy was lifted
in the long-wavelength limit $q \le 1/d$ though not without some
peculiarities. In particular, the order of the first four modes was
$\omega_{1+}<\omega_{1-}<\omega_{2-}<\omega_{2+}$. In bilayer
graphene the situation is more dramatic. From Eq.~(\ref{oddspec}) we
can see that, provided that $\ln{1/qa}\gg 1$, the frequencies of {\it many} odd modes decrease (soften)
considerably with increasing the wavelength (decreasing $q$) in
the region of intermediate wavelengths $q\sim 1/d$. However, that decrease in frequency does not persist at $q \to 0$ as the odd modes remain gapped there. The second unexpected finding is that the number
of nodes in the odd modes increases by 1 with increasing the
wavelength. This feature originates from the fast suppression of the
conductivity of the bilayer near $x=0$ essentially decoupling electric
currents in the $p$ and $n$ domains of the flake.

\section{Experimental implications}

The main obstacle in experimental observation of low-dimensional
plasmons is that the latter are typically gapless. This makes it
impossible to simply convert a photon into a plasmon with the
conservation of both energy and momentum, so that more complicated
experimental geometries are necessary.\cite{NB} In the case
considered in the present paper,  plasmons confined near graphene
bilayer {\it p-n} junctions are in general gapped, cf.\
Eqs.~(\ref{evenspec}) and (\ref{gaps}), and therefore straightforward
absorption of long-wavelength (infrared) radiation is possible. Since
the wavelength of the infrared radiation of interest to us reaches
millimeter range the corresponding electric field is virtually
homogeneous, so that only antisymmetric modes are
likely to be excited. In contrast, symmetric modes have electric
fields that are odd and therefore they do not couple to the
radiation polarized along the $x$ axis. This can be seen in Fig. \ref{electrostatics} where the electric field of the $(2+)$ mode along the $x$ axis in one half of the flake is the mirror reflection (along $x=0$) of the electric field in other half. In principle, symmetric
modes could be excited by the longitudinal (along the $y$ axis)
polarization. However, due to the conservation of momentum along the
$y$ axis and small values of momentum of infrared photons, the phase
space for such processes is rather restricted. Such absorption has
characteristic frequencies $\Delta =\sqrt{4eE_0/md}$. Using the
parameters of bilayer graphene we find
\begin{equation}
\label{signature} \Delta \approx 2.7~ \text{eV}\times
\frac{a_B}{l_E}\sqrt{\frac{a_B}{d}}.
\end{equation}
For electric fields, $\sim 10^6$ V/m, the value of electric length is
$l_E\sim 40$nm. Assuming $d\approx 1\mu$m for the size of the sample
we obtain from Eq.~(\ref{signature}) that $\Delta \sim 2$ meV. We
therefore expect a threshold signature in the absorption spectrum
of infrared radiation at these frequencies, which are very sensitive
to the magnitude of the electric field creating the {\it p-n}
junction.

We emphasize that the plasmon absorption happens {\it in addition}
to the intersubband electron-hole absorption. The latter, however,
is a smooth function of frequency. Indeed, interband absorption of
a photon with frequency $\omega$ is possible only as long as the
chemical potential (the distance to the Fermi level from the band
degeneracy point $p=0$) does not exceed $\hbar\omega/2$; see
Fig.~1(b). The former, expressed via the charge density,
Eq.~(\ref{eq30-1}), is simply $\pi |\rho|/2me=\pi E_0|x|/2me\sqrt{d^2-x^2}$.
Thus the corresponding transitions are allowed within a strip of
$|x|<d/\sqrt{1+(\pi E_0/me\omega)^2}$. The intensity of the
absorption is independent of frequency and simply determined by
the intersubband ac (``minimal") conductivity \cite{MC} of bilayer,
$e^2/2\hbar$. The frequency dependence of single-particle
absorption is therefore simply determined by the width of the
absorbing region,
\begin{equation}
\label{eh} A_{e-h}(\omega)\propto \frac{\omega
d}{\sqrt{\pi^2E^2_0+m^2e^2\omega^2}}.
\end{equation}
This expression describes a smooth background that exists in
addition to the usual intrasubband Drude absorption. Its
characteristic frequency $\hbar^2/ml_E^2$ depends linearly on the
applied electric field and should be easily distinguishable by
varying the electric field from the plasmon contribution whose frequency
(\ref{signature}) is proportional to $E^{1/2}_0$.

\section{Gated bilayer}

\begin{figure}[h]
\resizebox{.40\textwidth}{!}{\includegraphics{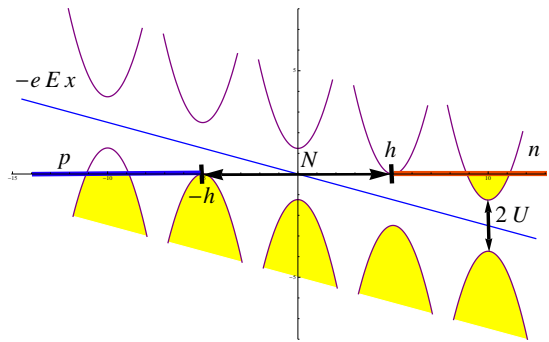}}
\caption{\label{fig2} (Color online) Electron band structure across the bilayer
with the energy gap of $2U$. The {\it p} (blue line starting from the point where the second lower peak on the left side touches the Fermi level and ending at the left end of the flake) and {\it
n} regions (red line on the right side of the axis) are separated by a neutral ({\it N}) strip (black
double-headed arrow) of width $2h=2U/eE$ determined by the energy gap and the slope
of the scalar potential.}
\end{figure}

The advantages of bilayer graphene for applications lie in the
possibility of inducing a bandgap with the interlayer bias, studied
both theoretically\cite{EMC,LAF,GLS,FMC} and
experimentally.\cite{KCM,ZTG,MLS} Since this is typically performed
via the nearby metallic gates, we now discuss how the properties of
guided plasmons are modified by the presence of such a gate, which
is assumed to be positioned a distance $D$ from the bilayer. In
addition, the presence of a bandgap leads to the appearance of a
neutral strip of width $2h$; see Fig.~\ref{fig2}.
In this section after explaining how the general
equations are changed, we consider three cases: (i) short
wavelengths, $q^{-1} \ll D$, (ii) intermediate wavelengths, $D \ll
q^{-1} \ll h$, and (iii) long wavelengths, $h \ll q^{-1} \ll d$.

Two modifications have to be made to our hydrodynamic equations. In particular, the induced electric field includes the additional contribution from image charges,
\begin{eqnarray}
\label{image} {\bf E}({\bf r},t) &=&{\bf E}_0-\nabla \int d^2r' \rho
({\bf r'},t) \nonumber\\&&\times\left( \frac{1}{|{\bf r}-{\bf r'}|}
-\frac{1}{\sqrt{({\bf r}-{\bf r'})^2+4D^2}}\right).
\end{eqnarray}
The second modification should incorporate the presence of a gap
$2U$ in the energy spectrum, $\epsilon(p)=\pm\sqrt{U^2+(p^2/2m)^2}$.
The relation between the local chemical potential and the induced
charge density now becomes
\begin{equation}
\label{eq10}  \mu=-\text{sgn}~(\rho) ~\sqrt{U^2+\left(\frac{\pi
\rho}{2me}\right)^2}.
\end{equation}

The existence of a gap in the spectrum means that separation of
electrons and holes is  possible only when the applied in-plane
electric field exceeds some value, $E_0\ge U/ed$. Even above this
threshold {\it p} and {\it n} regions are separated by a neutral
strip of width $2h=2U/eE$, as shown in Fig.~2. The mean distribution
of the induced charges $\rho_0(x)$ is determined by the Thomas-Fermi
equation
\begin{eqnarray}
\label{eq15} E_0x&+&\text{sgn}(x)\sqrt{\frac{U^2}{e^2}+\frac{\pi^2
a_B^2 }{4}\rho_0^2(x)}\nonumber\\ && + 2 \int_0^d dx' \rho_0(x')
\ln{\frac{x+x'}{|x-x'|}}=0.
\end{eqnarray}
The second term under the square root can (similarly to Sec.~II) be
neglected everywhere except very close to the flake's edges [more
precisely, as long as $a_B \rho_0(x)$ is less than either
$E_0x$ or $U/e$]. The resulting linear equation has been solved in
Ref.~\onlinecite{SMR} in relation to the polarization of a nanotube
array in the external electric field, yielding
\begin{equation}
\label{eq16} \rho_0(x)=\text{sgn}(x)~E_0
\sqrt{\frac{x^2-h^2}{d^2-x^2}}\:\Theta(|x|-h),
\end{equation}

The equation of motion for plasmon oscillations is also modified by
the  gap in the spectrum, as explained in Appendix A; see
Eq.~(\ref{withgap}). Solution of the corresponding general equation
for the oscillating density is beyond the scope of the present
paper. However, great simplification occurs if the gap is not very
small, so that the width of the neutral strip is much greater than
the Bohr radius, $h\gg a_B$. In this case the second term in the
denominator of Eq.~(\ref{withgap}) is negligible compared with $U$,
which amounts to the substitution
\begin{equation}
\label{to} |\rho_0 |\to \frac{\pi \rho_0^2}{2me U}
\end{equation}
in Eq.~(\ref{eq30}). The second change to Eq.~(\ref{eq30}) comes
from taking into account the image charges from Eq.~(\ref{image}).
This is done by replacing $K_0(q|x-x'|)$ with
\begin{eqnarray}
\label{kernel} K_0(q|x-x'|)-K_0[q\sqrt{(x-x')^2+4D^2}]\nonumber\\ =
\frac{1}{2}\int\limits_{-\infty}^\infty
\frac{dk}{\sqrt{q^2+k^2}}\left(1-e^{-2D\sqrt{q^2+k^2}}\right)e^{ik(x-x')}.
\end{eqnarray}
 Below we elucidate plasmon dispersion for the case of  a wide flake and a gate situated in
close proximity to it, $D\ll h \ll d$.

(i) Short wavelengths, $q^{-1} \ll D$. In this case the effect of
the image charges is negligible. Plasmons propagate along a boundary
between a charged ({\it p} or {\it n}) domain and a neutral strip
({\it N}) and decay over a distance short compared to $h$. Thus
modes localized near the {\it p-N} boundary are exponentially weakly
coupled to {\it n-N} modes. The corresponding plasmon frequencies,
\begin{equation}
\omega^2 \sim \frac{E_0^2 h}{m^2d^2 U},
\end{equation}
are therefore doubly degenerate and independent of the wavelength.

(ii) Intermediate wavelengths, $D \ll q^{-1} \ll h$. The wavelength
is still short enough  to ensure that plasmons propagating near the
two edges of the neutral strip do not ``talk,'' but the Coulomb
interaction is screened by the gate. The expression in Eq.~(\ref{kernel}) can be approximated with $2\pi D\delta(x-x')$. Since
only small $\tilde x=x-h \ll h$ are of interest, the density in
Eqs.~(\ref{eq15}) and (\ref{to}) can be approximated with ($\tilde
x>0$)
\begin{equation}
\rho_0^2(\tilde x) =\frac{2hE_0^2}{d^2}\tilde x
\end{equation}
By virtue of the $\delta$ function the equation for the oscillating
charge density becomes a differential one,
\begin{align}
\label{ch6eq39} \frac {1}{\tilde{x}}\frac{d}{d\tilde{x}}\left(
\tilde{x} \frac {d\chi(\tilde{x})}{d \tilde{x}} \right)
+\left(\frac{\omega^2}{C^2\tilde{x}}-q^2\right)\chi(\tilde{x}) =0,
\end{align}
where $C^2=\frac{4 \pi^2 E_0^2 h D}{d^2 m^2 U}$. This equation is
identical to that of a two-dimensional ``hydrogen atom'' with $-q^2$
playing the role of the energy and $\omega^2/C^2$ the role of the
interaction constant. The corresponding quantization condition is
simply $q=\omega^2/2C^2(n+1/2)$, which yields the spectrum
\begin{equation}
\label{ch6eq45} \omega^2_n=  (n+1/2) \frac{8 \pi^2 \,E_0^2 h
\,D}{d^2 m^2 U} q, ~~~n=0,1,2,\,.\,.\,.\,.
\end{equation}
The corresponding eigenfunctions are given by the confluent
hypergeometric function of the first kind,
\begin{equation}
\label{ch6eq46} \chi_n(\tilde{x})=  e^{-q\tilde x}
~_1F_1(-n,1,2q\tilde x).
\end{equation}
Remarkably, the solutions (\ref{ch6eq46}) do not vanish at $\tilde x
= x-h =0$. This indicates that the shape of the boundary of the
neutral strip, created by ``soft'' electrostatic confinement, does not
remain linear and itself fluctuates due to plasmon excitations.

(iii) For yet longer wavelengths, $h \ll q^{-1}\ll d$, the presence
of the neutral strip is irrelevant as the field of plasmons extends
over much larger distances. Assuming that the wavelength is still
much smaller than the width $d$, we can use $\rho^2=E_0^2x^2/d^2$,
cf. ~Eq.~(\ref{eq30-1}). The $\delta$ function approximation from the
previous paragraph still holds, yielding
\begin{equation}
\label{ch6eq48}
\frac {1}{\xi^2}\frac{d}{d\xi}\left(
\xi^2 \frac {d\chi} {d \xi} \right)
+\left(\frac{\omega^2}{C^2_1\xi^2}-1\right)\chi =0,
\end{equation}
where $\xi=qx$, and $C_1^2=\frac{2\pi^2 E_0^2 D}{m^2 d^2 U}$. With the help of the substitution, $\chi(\xi)=f(\xi)/|\xi|$, Eq.~(\ref{ch6eq48}) is reduced to
\begin{equation}
\label{WKB} \frac{d^2f}{d\xi^2}+
\left(\frac{\omega^2}{C^2_1\xi^2}-1\right)f=0,
\end{equation}
which is the one-dimensional Schr\"odinger equation for the potential $\propto -1/\xi^2$. The singularity at $\xi=0$ needs to be regularized by cutting off at $\xi\approx qa$. Using the Bohr-Sommerfeld condition,
\begin{equation}
\label{ch6eq49} 4\int\limits_{qa}^{\omega/C_1} d\xi \sqrt{\frac{\omega^2}{C_1^2\xi^2}-1}=2\pi \left(n+\frac{1}{2}\right),
\end{equation}
we find the spectrum to be
\begin{equation}
\label{ch6eq50}
\omega^2 = \frac{\pi^4E_0^2 D }{2m^2d^2 U\ln^2(qa)}(n+1/2)^2, ~~~n=0,1,2,\,.\,.\,.\,.
\end{equation}
The result (\ref{ch6eq50}) becomes more accurate for $n\gg 1$, since
the WKB applicability  condition for Eq.~(\ref{WKB}) requires that
$\omega \gg C_1$. Still, even for $n\sim 1$ the expression
(\ref{ch6eq50}) captures the correct dependence on the gap $U$ and
the strength of the external field $E_0$.

To conclude the discussion of the gated bilayer, let us note that
only in the intermediate case (ii) does the plasmon frequency
demonstrate dependence on the wave number, namely when the
wavelength is long enough to ensure screening by image charges, but
still sufficiently short compared with the width of the neutral
strip. In all three regimes the dependence of plasmon spectra on
electric field $E_0$ is linear in contrast to a gapless case, where
the corresponding dependence is square root; see
Eqs.~(\ref{evenspec}), (\ref{oddspec}), and (\ref{gaps}).

\section{Summary}

Bilayer graphene is a gapless (or weakly gapped) system that is
charge neutral when undoped. Separation of charges, however,
occurs with even weak external electric fields applied along the
plane of bilayer; see Fig.~1. The induced charge density
Eq.~(\ref{eq30-1}) follows from the solution of the electrostatic
problem. The gradient of charge density near the {\it p}-{\it n}
junction line $x=0$ leads to the confinement of charge
oscillations (plasmons) resulting in their one-dimensional
propagation along the junction. Physically, confinement in the
transverse direction can be understood as follows. Since plasmon
``stiffness" increases with increasing the local charge density,
plasmons of lower frequency tend to be located closer to the {\it
p}-{\it n} junction and to decay into the region of higher density.

Within a continuous hydrodynamic approach for charge-density
oscillations the plasmon modes are determined from the
integrodifferential eigenvalue problem, Eq.~(\ref{eq53}). The
latter allows for exact solutions, Eqs.~(\ref{evensol}) and
(\ref{oddsol}), with the discretization of the spectrum being a
consequence of the regularization of the logarithmic singularity,
Eq.~(\ref{dispersion}). For small wave vectors $q$ the lowest mode is
an even solution that is a conventional 1D plasmon with logarithmic
velocity. At $q\sim 1/d$ the order is reversed and the first odd
solution becomes the mode with the lowest energy.

From the practical standpoint guided plasmon modes could
potentially be used for ``plasmon transistors.''\cite{HAA} The
immediate experimental signatures of the guided plasmons, however,
can be most directly measured in the optical absorption. While the
electron-hole intersubband excitations lead to a smooth
continuum, Eq.~(\ref{eh}), the gapped plasmon spectrum of
antisymmetric modes should result in a threshold behavior at a frequency
$\sim eE_0/md$ sensitive to the applied electric field.

\acknowledgments

Useful discussions with M. E. Raikh are gratefully acknowledged. The
work was supported by the Department of Energy, Office of Basic
Energy Sciences, Grant No.~DE-FG02-06ER46313, and by the Research
Corporation for Science Advancement.

\appendix

\section{Hydrodynamic equations}

Equations (\ref{eq18}) and (\ref{eq17}) can be derived from the
Boltzmann equation for the electron distribution function $f_{\bf
p}$,
\begin{equation}
\label{boltz} \frac{\partial f_{\bf p}}{\partial t} +{\bf v}\cdot
\nabla f_{\bf p}+e{\bf E}\cdot \frac{\partial f_{\bf p}}{\partial
\bf p}=I[f_{\bf p}],
\end{equation}
where the right-hand side is the collision integral.\cite{LLX}
Since the latter conserves the number of electrons, integrating the
equation over the entire momentum space yields the continuity
equation (\ref{eq18}). Hydrodynamic approximation is applicable
when a distribution function deviates only slightly from a local
equilibrium distribution with the chemical potential $\mu({\bf
r},t)$ (which is equivalent to specifying the local density
$\rho$). The deviation is due to the drift of particles with the
average velocity being considerably smaller than the Fermi
velocity $v_F$, and characterized by a local current density ${\bf
J}({\bf r},t)$:
\begin{equation}
\label{anzats} f_{\bf p}({\bf r},t)=\Theta (\mu({\bf
r},t)-\epsilon_p) +2\frac{{\bf v}\cdot {\bf J}({\bf r},t) }{ev_F^2
\nu(\mu)} \delta [\mu({\bf r},t)-\epsilon_p],
\end{equation}
where $\nu(\mu)$ is the density of states at the Fermi level. Such
an ansatz is nothing but the expansion over angular harmonics
truncated after the first term.  Multiplying Eq.~(\ref{boltz}) by
$e{\bf v}$ and performing the same operation, we can obtain the
equation for electric current,
\begin{eqnarray}
\label{cur} \frac{\partial \bf J}{\partial t}+e\sum_{\bf p} {\bf
v}({\bf v}\cdot \nabla f_{\bf p})+e^2\sum_{\bf p} {\bf v}\biggl({\bf
E}\cdot \frac{\partial f_{\bf p}}{\partial \bf
p}\biggr)\nonumber\\=e\sum_{\bf p} {\bf v}I[f_{\bf p}].
\end{eqnarray}
The last term represents collision relaxation of the electric
current and for small currents (linear response) may be written as
$-{\bf J}/\tau$, where $\tau$ is the transport mean free time.
Substituting Eq.~(\ref{anzats}) into Eq.~(\ref{cur}) we get
\begin{equation}
\frac{\partial \bf J}{\partial t}+\frac{{\bf
J}}{\tau}=\frac{1}{2}ev_F^2\nu(\mu)(e{\bf E}-\nabla \mu).
\end{equation}
This expression generalizes the usual Drude conductivity onto the
case of coordinate- and time-dependent density of 2D electron gas.
For frequencies exceeding the collision rate the second term in
the left-hand side can be neglected.  If the gap is present the
spectrum is $\epsilon(p)=\sqrt{U^2+(p^2/2m)^2}$. Calculating the
Fermi velocity $v_F=d\mu/dp_F$ and the density of states,
$\nu(\mu)=2p_F/\pi v_F$, we obtain
\begin{equation}
\label{withgap} \dot{\bf
J}=\frac{\pi}{2m^2e}\frac{\rho^2}{\sqrt{U^2+\left(\frac{\pi
\rho}{2me}\right)^2}}(e{\bf E}-\nabla\mu).
\end{equation}
In the case where there is no gap, $U=0$, Eq.~(\ref{eq17}) is recovered.

\section{Integral relations}

Integral relations, Eqs. (\ref{intrel1}) and (\ref{intrel2}), can be
established by considering the integral
\begin{eqnarray}
\label{expint} {P}\!\! \int\limits_{-\infty}^\infty \!dt\,
\frac{\cosh t\, e^{i\alpha t}}{\sinh t-\sinh \tau}.
\end{eqnarray}
This integral can be evaluated by noticing that in the upper plane
of the complex variable $t$, the integrand is exponentially
suppressed on the infinite semicircle, $t=R e^{i\varphi}$, $0\leq
\varphi\leq\pi$, $R\rightarrow\infty$; see Fig.~6(a). Then the
integral equals the sum of residues at $t=\tau+2\pi n i$ and
$t=-\tau+\pi(2n-1)i$, where $n=1,2,\,.\,.\,.$, times $2\pi i$, plus the residue
at $t=\tau$ times $\pi i$, which is due to the principal
integration. After the integral of Eq. (\ref{expint}) is evaluated,
Eqs. (\ref{intrel1}) and (\ref{intrel2}) hold as being the real and
imaginary parts of it, respectively.

Upon replacing $k\mapsto\sinh \tau$ and $k^\prime\mapsto\sinh t$,
the integrodifferential equation (\ref{eq53}) acquires the
form
\begin{eqnarray}
\label{refinteq} \omega^2 \tilde{\chi}(\tau)=-\frac{4\pi e E}{md}
~{P}\!\! \int\limits_{-\infty}^\infty \!\frac{dt}{2\pi}\,
\frac{\cosh t}{\sinh t-\sinh \tau}\frac{d\tilde{\chi}(t)}{dt},
\end{eqnarray}
where $\tilde{\chi}(t)=\chi(\sinh t)$. Equations
(\ref{intrel1}) and (\ref{intrel2}) imply that
$\tilde{\chi}(t)=\cos(\alpha t)$ and $\tilde{\chi}(t)=\sin(\alpha
t)$ solve the integrodifferential equation, Eq. (\ref{refinteq}),
yielding the spectra Eqs. (\ref{evenspec}) and (\ref{oddspec}),
respectively. Going back to the original variable,
$t=\text{arcsinh}\, k$, we recover the even and odd solutions, Eqs.
(\ref{evensol}) and (\ref{oddsol}), which in addition are multiplied
by factors $\cosh(\pi\alpha/2)$ and $i\sinh(\pi\alpha/2)$,
respectively, for the sake of further convenience. This can be
done without destroying the solutions as these factors are
constants with respect to the variable $k$.
\begin{figure}[b]
\resizebox{.25\textwidth}{!}{\includegraphics{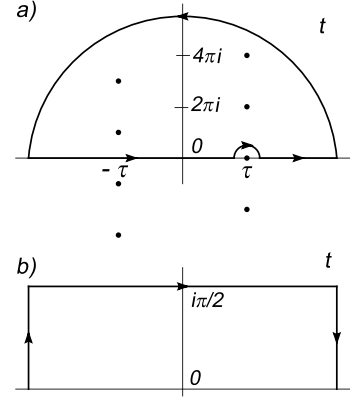}}
\caption{\label{fig5} (a) Contour in the complex plane for the
calculation of the integral (\ref{expint}); the singularities in the
integrand are indicated by the dots; (b) deformation of the contour
from the real axis for the calculation of the integral
(\ref{rhox2}). }
\end{figure}

In the remaining portion of this Appendix we establish the relation Eq.
(\ref{Fourier}). Consider the Fourier transform
\begin{eqnarray}
\label{Ft} \chi^{(+)}_\alpha(\xi)=\int\limits_{-\infty}^\infty
\frac{dk}{2\pi}\, e^{ik\xi} \chi^{(+)}_\alpha (k).
\end{eqnarray}

After performing a change, $k= \sinh t$, and integrating by parts,
we rewrite Eq. (\ref{Ft}) in the form
\begin{eqnarray}
\label{rhox2}
\chi^{(+)}_\alpha(\xi)=\frac{\alpha\cosh(\pi\alpha/2)}{2\pi i\xi}
\int\limits_{-\infty}^\infty \!dt\, e^{i\xi\sinh t} \sin(\alpha
t).
\end{eqnarray}
The integrand in Eq.~(\ref{rhox2}) is a holomorphic function of the
complex variable $t$, so that the integration contour can be
deformed from real axis to the horizontal line, $-\infty+i\pi/2\leq
t\leq\infty+i\pi/2$; see Fig.~6(b). In doing this we also notice that
the integrals over the vertical parts, $-\infty\leq
t\leq-\infty+i\pi/2$ and $\infty+i\pi/2\leq t\leq\infty$, are
suppressed. Then Eq. (\ref{rhox2}) turns into the relation
\begin{eqnarray}
\label{rhox3}
\chi^{(+)}_\alpha(\xi)=&&\!\!\!\!\!\frac{\alpha\cosh(\pi\alpha/2)}{2\pi
i\xi}\!\! \int\limits_{-\infty}^\infty\!\! dt\, e^{-\xi\cosh
t}\bigl[\sin(\alpha
t)\cosh(\pi\alpha/2)\nonumber\\
&&+i\cos(\alpha t)\sinh(\pi\alpha/2)\bigr].
\end{eqnarray}
The first term in the rectangular brackets turns to zero as being
an odd function of $t$, and we finally arrive at the integral
representation,
\begin{eqnarray}
\label{rhox4} \chi^{(+)}_\alpha(\xi)=\frac{\alpha\sinh(\pi\alpha)}
{2\pi \xi} \int\limits_{0}^\infty \!dt\, e^{-\xi\cosh t}
\cos(\alpha t).
\end{eqnarray}

\end{document}